\documentclass{emulateapj}

\newcommand{\Kepler}{{\it Kepler}}

\newcommand{\betapicb}{$\beta$ Pictoris b}
\newcommand{\betapic}{$\beta$ Pictoris}

\newcommand{\be}{\begin{equation}}
\newcommand{\ee}{\end{equation}}

\newcommand{\msun}{M$_\odot$}
\newcommand{\rsun}{R$_\odot$}
\newcommand{\mj}{M$_J$}
\newcommand{\rj}{R$_J$}

\newcommand{\kms}{\ensuremath{\rm km\,s^{-1}}}
\newcommand{\ms}{\ensuremath{\rm m\,s^{-1}}}
\newcommand{\cms}{\ensuremath{\rm cm\,s^{-1}}}
\newcommand{\ron}{\color{black}}

\newcommand{\thisplanet}{\Kepler-1625b}
\newcommand{\thismoon}{\Kepler-1625b I}

\usepackage{xcolor}
\usepackage{hyperref}
\usepackage{breakurl}
\usepackage{amsmath}
\DeclareMathOperator{\sign}{sign}
\usepackage{graphicx}
\usepackage{verbatim}
\usepackage{booktabs}
\usepackage{wasysym}

\slugcomment{}

\shorttitle{Radial Velocity Detection of Exomoons}
\shortauthors{Vanderburg, Rappaport, and Mayo}

\begin{document}


\title{Detecting Exomoons Via Doppler Monitoring of Directly Imaged Exoplanets}
\author{Andrew Vanderburg\altaffilmark{1,$\dagger$}, Saul A. Rappaport\altaffilmark{2}, and Andrew W. Mayo\altaffilmark{3,4}}

\altaffiltext{1}{Department of Astronomy, The University of Texas at Austin, Austin, TX 78712, USA}
\altaffiltext{2}{Department of Physics, and Kavli Institute for Astrophysics and Space Research, M.I.T., Cambridge, MA 02139, USA}
\altaffiltext{3}{DTU Space, National Space Institute, Technical University of Denmark, Elektrovej 327, DK-2800 Lyngby, Denmark}
\altaffiltext{4}{Centre for Star and Planet Formation, Natural History Museum of Denmark \& Niels Bohr Institute, University of Copenhagen, \O ster Voldgade 5-7, DK-1350 Copenhagen K.}

\altaffiltext{$\dagger$}{NASA Sagan Fellow, \url{avanderburg@utexas.edu}}


\begin{abstract}
Recently, Teachey, Kipping, and Schmitt (2018) reported the detection of a candidate exomoon, tentatively designated \thismoon, around a giant planet in the \Kepler\ field. The candidate exomoon would be about the size and mass of Neptune, considerably larger than any moon in our Solar System, and if confirmed, would be the first in a new class of giant moons or binary planets. Motivated by the large mass ratio in the \thisplanet\ planet and satellite system, we investigate the detectability of similarly massive exomoons around directly imaged exoplanets via Doppler spectroscopy. The candidate moon around \thisplanet\ would induce a radial velocity signal of about 200 \ms\ on its host planet, large enough that similar moons around directly imaged planets orbiting bright, nearby stars might be detected with current or next generation instrumentation. In addition to searching for exomoons, a radial velocity survey of directly imaged planets could reveal the orientations of the planets' spin axes, making it possible to identify Uranus analogs.

\end{abstract}


\section{Introduction}

Exomoons, or moons orbiting planets around stars other than our own Sun, are hard to detect. It is a challenge, in fact, to even detect the extrasolar planets which might host these moons -- only in the last three decades has astronomical instrumentation advanced to the point where exoplanets could be confidently claimed \citep{campbellwalker, latham, wolszczan, mayor}. Exoplanets are often detected by measuring small signals in light coming from their host stars -- such as low-amplitude periodic modulations in the radial velocity (RV) of the host star \citep[e.g.][]{mayor}, small dimmings in the host star's brightness as the planet transits, passing in front of the star in our line of sight \citep[e.g.][]{charbonneau}, or perturbations to the gravitational potential in a system from a planet, illuminated through gravitational microlensing \citep[e.g.][]{bondmicrolensing}. In rare cases, it is possible to actually take images (and spectra) of a planet by angularly resolving the planet and the star using adaptive optics (AO) imaging \citep[e.g.][]{marios}. Advances in adaptive optics and the next generation of extremely large telescopes promise to make these directly imaged planets more common \citep[e.g.][]{eltao}. 

Detecting exomoons orbiting these planets adds an extra level of difficulty because, in general, moons contribute {\ron only tiny} perturbations on top of the already small signals caused by exoplanets. So far, there are numerous proposed methods for detecting exomoons, ranging from measuring photocenter motion of a directly imaged but unresolved planet/moon system \citep{cabreraschneider, agol2015}, to detecting transits of self-luminous planets by their moons\footnote{If a moon is found to transit its planet, even more could be learned by observing the Rossiter McLaughlin effect as the moon transits its host planet \citep{helleralbrecht}.} \citep{cabreraschneider}, to observing changes in the polarization signature of giant exoplanets \citep{senguptamarley}. Today, however, the most sensitive searches for exomoons look for small perturbations to transit light curves from stars hosting giant transiting planets \citep[e.g.][]{kipping22, kipping41,hippke, teacheykipping}. A moon orbiting a transiting planet would cause small variations in the timing and duration of transits of the planet across the stellar host \citep{Sartoretti, kipping2009}. Moons would also block starlight themselves, perturbing the shape of transits and potentially increasing the total transit duration. It is possible to identify exomoon candidates by detecting changes to the average shape of the host planets' transits \citep{hellerose}, but the most sophisticated searches for exomoons today involve full photo-dynamical modeling of light curves of stars hosting transiting planets using specialized codes \citep{kippinghek}. 

\begin{deluxetable*}{lccccc}[ht!]
\tablewidth{0pt}
\tablecaption{Summary of Relevant Radial Velocity Signals}

\tablehead{
 \colhead{Section} & \colhead{Signal } & \colhead{Self-Luminous} & \colhead{Reflected-Light} & \colhead{RV}& \colhead{Timescale}\\
\colhead{ } & \colhead{ } & \colhead{Planets} & \colhead{Planets} & \colhead{Amplitude}& \colhead{ } }
\startdata
\ref{reflex} & Exomoon Reflex Motion & Yes & Yes & Up to $\sim$ 1 \kms& days to months   \\
\ref{orbit} & Planet Orbit & Yes & Yes & $\sim 5-50$ \kms & years to decades \\
\ref{illumination} & Planetary Illumination & No & Yes & Up to $\sim$ 10 \kms& same as planet's orbit  \\
\ref{activity} & Planetary Activity & Yes & Yes (lower amplitude) & Up to $\sim $100 \ms & hours to days \\
\ref{peakpulling} &  Peak-pulling by Exomoon & Maybe & Yes & similar to moon's & fraction of moon's orbit  \\
& & & & orbital motion & near conjunction phases  \\
\ref{diskclumps} & Disk Clump Occultation & Maybe & No & Up to $\sim$ 100 \ms & Aperiodic variations\\
& & & &  & over days to weeks   \\

\enddata
\label{signalsummary}
\end{deluxetable*}
Recently, \citet{teacheykipping} reported {\ron perhaps} the strongest yet exomoon candidate around a giant planet in the \Kepler\ field, \thisplanet\footnote{\thisplanet\ was identified as a transiting planet candidate by the \Kepler\ pipeline \citep{mullally} and subsequently statistically validated by \citet{morton16}.}. A photodynamical fit to three transits of the planet observed by \Kepler\ during its original mission suggests that the candidate moon, \thismoon, is about the size and mass of Neptune. While the detection is tentative, and historically, exomoon candidates have not survived further scrutiny \citep{cabrera, kipping90g}, the candidate around \thisplanet\ has so far passed standard tests including cross-validation between different data segments and careful inspection of pixel-level \Kepler\ data. The inferred orbit of the moon is physically plausible, far enough from the planet to avoid Roche lobe overflow, and close enough to remain dynamically stable. Furthermore, the inferred masses of both the moon and planet from the photodynamical fit are consistent with empirically determined mass/radius relationships. The candidate signal was therefore compelling enough to warrant follow-up observations with the {\em Hubble Space Telescope}, which were executed in October 2017, and which may either confirm or refute the proposed moon scenario.  

If the candidate exomoon around \thisplanet\ is confirmed by {\em Hubble} observations, it would be a historic discovery. In addition to being the first moon (or binary planet, depending on one's definition) discovered around a planet outside our solar system, it would be the first in a new class of massive moons \citep{hellerkepler1625}, larger than most of the {\em planets} known to exist in our galaxy \citep{fressin}.  While {\em a priori}, the discovery of an object so different from anything found in our solar system might seem unlikely, historically exoplanet searches have yielded one surprising result after another, from hot Jupiters \citep{mayor} to multi-planet systems orbiting pulsars \citep{wolszczan}. These discoveries have shown that searches in new regimes, even for objects with no known analogs within our own solar system, can bear fruit \citep{struve}. 

Here, we propose that massive exomoons similar to \thismoon\ might be detected by radial velocity monitoring of directly imaged planets. Recently, significant progress has been made in detecting thermal light from hot exoplanets using high-resolution near infrared spectrographs \citep{snellen, bryan, schwarz, Hoeijmakers}. The detection of spectral features at high resolution makes it possible to measure the radial velocity of the {\em planets} themselves. While others have previously discussed detecting exomoons with radial velocity observations of directly imaged planets {\ron \citep[][]{hook, cabreraschneider, hellertransits, lillobox}}, here we quantitatively assess the various RV signals one might expect to detect when observing directly imaged planets. This paper is organized as follows. In Section \ref{signals} we describe and estimate the RV signals induced by exomoons on their host planets as well as other ``nuisance'' signals due to the planets' orbital motion, changes to the planets' illumination, and inhomogeneities on the planets' surfaces. We find that the signals of giant, \thismoon-like exomoons should either be easily separable from or dominate over nuisance signals, making their detection astrophysically plausible. In Section \ref{detectability}, we estimate the feasibility of measuring radial velocities with high enough precision to detect RV signals caused by massive orbiting exomoons and conclude that massive moons could be detected around bright directly imaged planets with present-day instrumentation\footnote{In particular, we show that detecting massive exomoons could be feasible using a high-resolution infrared echelle spectrograph behind a modern high-contrast adaptive optics system on an 8m class telescope. While there is not yet such a spectrograph/AO combination, the major technical hurdles necessary for the observations have been overcome.}. In Section \ref{discussion}, we discuss the potential impact of a radial velocity exomoon search and point out that a Doppler monitoring program aimed at detecting exomoons would naturally yield interesting additional results even if no exomoons are detected. Finally, in Section \ref{conclusions}, we summarize our results and their implications. 

\section{Relevant Radial Velocity signals} \label{signals}

\begin{figure*}[ht!] 
   \centering
   \includegraphics[width=6.5in]{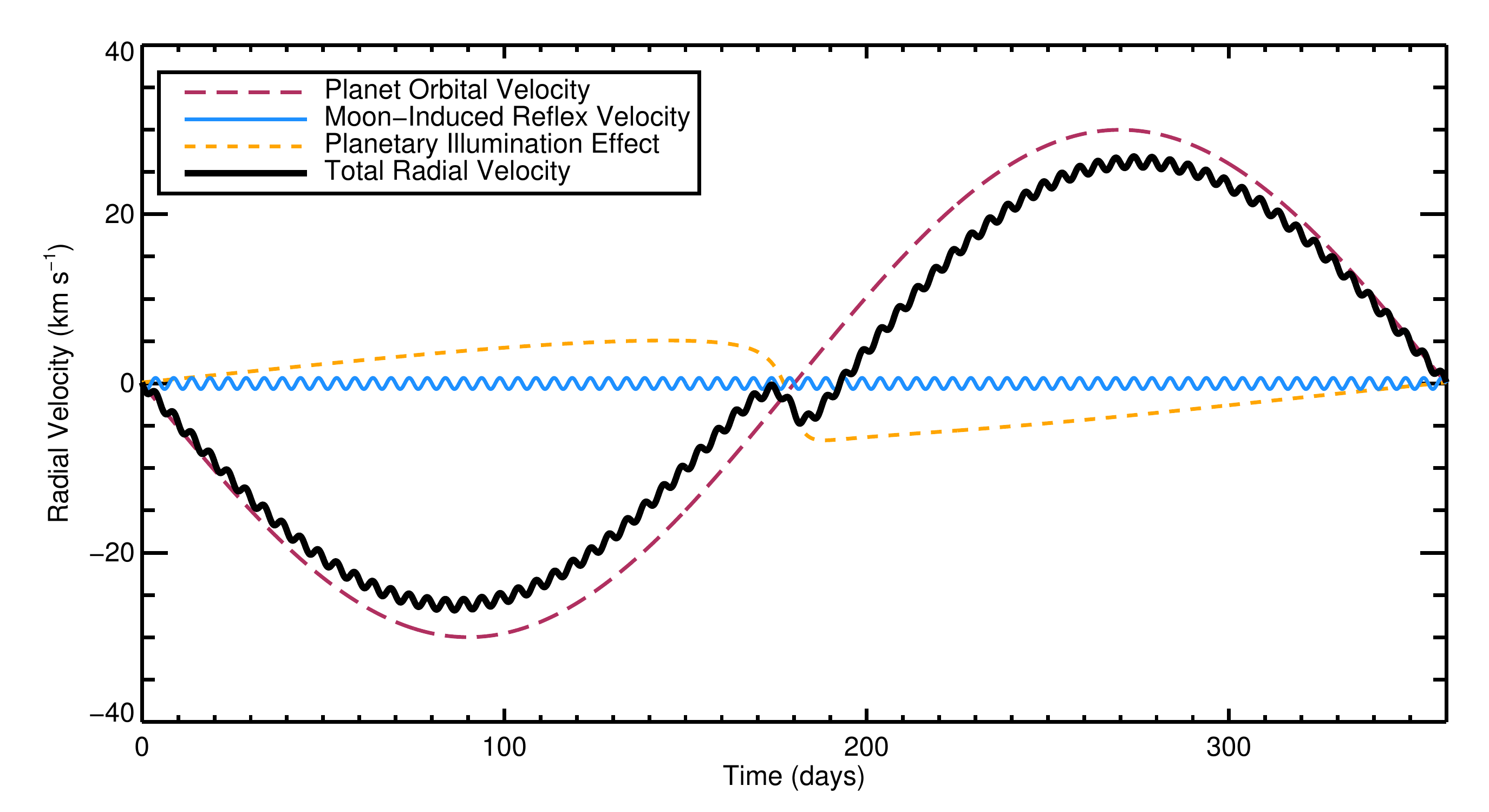} 
   \caption{Simulated RV signals of a hypothetical directly imaged exoplanet illuminated by reflected light from its host star. The planet is the mass and radius of Jupiter, and has a rotation period of 18 hours. The various components of the RV signal are shown as thin colored lines, and the total RV signal for the planet is shown as a thick black line. The planet orbits a sun-like star in a 360 day period and hosts a Neptune-mass exomoon in a 5 day orbit. The RV contribution due to the varying illumination of the planet is calculated for a planet with an orbital inclination of 85$^\circ$, with the spin angle $\xi = $60$^\circ$. The planetary activity contribution for this particular planet, assuming photometric variations similar to Neptune, is smaller than the width of the curves.}
   \label{signalfigure}
\end{figure*}

In this section we discuss the various radial velocity signals which might be present in observations of directly imaged giant planets. We summarize the different signals in Table \ref{signalsummary} and show an example of how these signals might present for a moon-hosting exoplanet imaged in reflected light in Figure \ref{signalfigure}. 

\subsection{Reflex Motion from an Exomoon}\label{reflex}

A moon orbiting an exoplanet will induce a Keplerian radial velocity reflex motion on the planet. The radial velocity semiamplitude of the Keplerian signal, $K_{p}$, induced by a moon with mass $m_{\leftmoon}$ on a planet with mass $m_{p}$ is given by: 
{\ron
\begin{equation}\label{planetsemiamplitude}
    K_{p} = \left[ \frac{2 \pi G}{P_{\leftmoon}(1-e_{\leftmoon}^2)^{3/2}} \frac{m_{\leftmoon}^3\sin^3{i_{\leftmoon}}}{(m_p + m_{\leftmoon})^2}\right]^{1/3}
\end{equation}
}

\noindent where $G$ is Newton's gravitational constant, $e_{\leftmoon}$ is the orbital eccentricity, and $i_{\leftmoon}$ is the orbital inclination. If we assume circular orbits and take the planet's mass to be much larger than the moon's mass ($m_p >> m_{\leftmoon}$), the expression for the RV semiamplitude reduces to: 

\begin{equation}
    K_{p} \approx \left[ \frac{2 \pi G}{P m_p^2}\right]^{1/3} m_{\leftmoon}\sin{i_{\leftmoon}}
\end{equation}

\begin{deluxetable}{lcccc}
\tablewidth{0pt}
\tablecaption{Largest RV Signals From Solar System Moons}

\tablehead{
\colhead{ } & \colhead{Moon} & \colhead{Planet}& \colhead{Orbital } &\colhead{RV Semi-}\\
\colhead{ } & \colhead{Name} & \colhead{Name}& \colhead{ Period (days)} &\colhead{Amplitude (\ms)}}
\startdata
1. & Charon & Pluto & 6.39 & 21.91 \\
2. & Moon & Earth & 27.32 & 12.46 \\
3. & Titan & Saturn & 15.95 & 1.32 \\
4. & Triton & Neptune & 5.88 & 0.92 \\
5. & Ganymede & Jupiter & 7.16 & 0.85 \\
6. & Io & Jupiter & 1.77 & 0.82 \\
7. & Callisto & Jupiter & 16.69 & 0.46 \\
8. & Europa & Jupiter & 3.55 & 0.35 \\
9. & Titania & Uranus & 8.71 & 0.14 \\
10. & Oberon & Uranus & 13.46 & 0.10 \\
\enddata
\tablecomments{Moon data are taken from \url{https://ssd.jpl.nasa.gov/?sat_phys_par} and \url{https://ssd.jpl.nasa.gov/?sat_elem}.} 
\label{moonsemiamplitudes}
\end{deluxetable}

Table \ref{moonsemiamplitudes} lists the Solar System moons which induce the largest RV semiamplitude on their host planet. While most moons in the Solar System cause only low-amplitude radial velocity reflex motion on their host planets, a handful of the more massive moons (in particular, Charon and Earth's moon) can induce signals of tens of meters per second. 

If real, the candidate exomoon around \thisplanet\ would cause a far larger RV signal on its host planet than the small moons in our Solar system. A Neptune-mass moon orbiting a 10 \mj\ planet with  a 1.8 day orbital period\footnote{The 1.8 day orbital period is estimated from the reported semimajor axis and planet mass from the \citet{teacheykipping} photodynamical fits.} like \thismoon\ should induce a radial velocity signal on its host planet of $K_{p} \approx $ 200 \ms. While the RV semi-amplitude induced by \thismoon\ is large, it is by no means the extreme. If this same moon were found orbiting a planet with the mass of Jupiter, it could induce an RV semiamplitude as large as 900 \ms, and if the moon were orbiting near its Roche limit ($P\approx 10$ hours), the RV semiamplitude could be boosted by another 50\%. {\ron \thismoon-like moons are not the only class which might induce large RV semiamplitudes; an Earth-mass exomoon in a similar orbit around a Jupiter-mass planet would induce a signal with an amplitude of approximately 50 \ms, while a rocky super-Earth-sized moon \citep[analogous to super-Earth planets like LHS 1140 b,][]{dittmann} in a 2 day orbit around a Jupiter-mass planet could induce a signal upwards of 300 \ms. In favorable conditions, rocky exomoons like these might be habitable \citep{hill2018}.}

Thanks to advances in instrumentation and data processing techniques \citep{butler, harps}, in the last few decades it has become possible to detect RV signals with semiamplitudes as small as about 1 \ms\ \citep[e.g.][]{hd219134} around stars brighter than about V=10$^{\rm th}$ magnitude, and future instruments are being designed with the goal of remaining stable at the level of 10 \cms. While achieving RV precision of a few meters per second or better in spectroscopic observations of directly imaged exoplanets is likely out of reach, instrumental stability should not be a limiting factor in RV searches for exomoons, which can induce RV variations considerably larger than those routinely detected with precise Doppler spectroscopy.

The orbital periods of Solar System moons range widely from the 7 hour orbits of Naiad about Neptune and Metis around Jupiter to the Neso's 27 year orbit around Neptune, but the most massive moons tend to have a more narrow range of orbital periods. Among the 17 Solar system moons with mass greater than $10^{20}$ kg, the orbital periods range from about 1.4 days to 80 days.  The estimated 1.8 day orbital period of \thismoon\ also falls within this range. 

\subsection{Orbit about the Host Star} \label{orbit}

{\ron The short-period radial velocity signal of an exomoon induced on its host planet will be overlaid on the longer-period Keplerian radial velocity signal of the planets' own orbit around its host star.  A planet's radial velocity signal from its orbit about a star with mass $M_\star$ in an orbit with period $P_o$ has a semi-amplitude $K_{o}$ given by:} 

\begin{equation}\label{outerorbitsemiamplitude}
     K_{o} \approx \left[ \frac{2 \pi G M_\star}{P_o}\right]^{1/3} \sin{i_o} 
\end{equation}

\noindent where $i_o$ is the inclination of the planet's orbit about the host star. The amplitude of this signal is large; for \thisplanet, the semiamplitude should be about 33 \kms, and for the directly imaged planet $\beta$ Pictoris b\footnote{Using orbital parameters from \citet{wang}.}, the semiamplitude should be about 13 \kms. Because this signal is strictly periodic and on a much longer period than any moon signals, it will be straightforward to model away the long-term orbital signal to detect short-period moon signals. The typical orbital periods for directly imaged planets will be years to decades, far longer than the typical moon orbital periods of days to weeks. 

\subsection{Planetary Illumination Effect}\label{illumination}

\begin{figure}[t!] 
   \centering
   \includegraphics[width=3.25in]{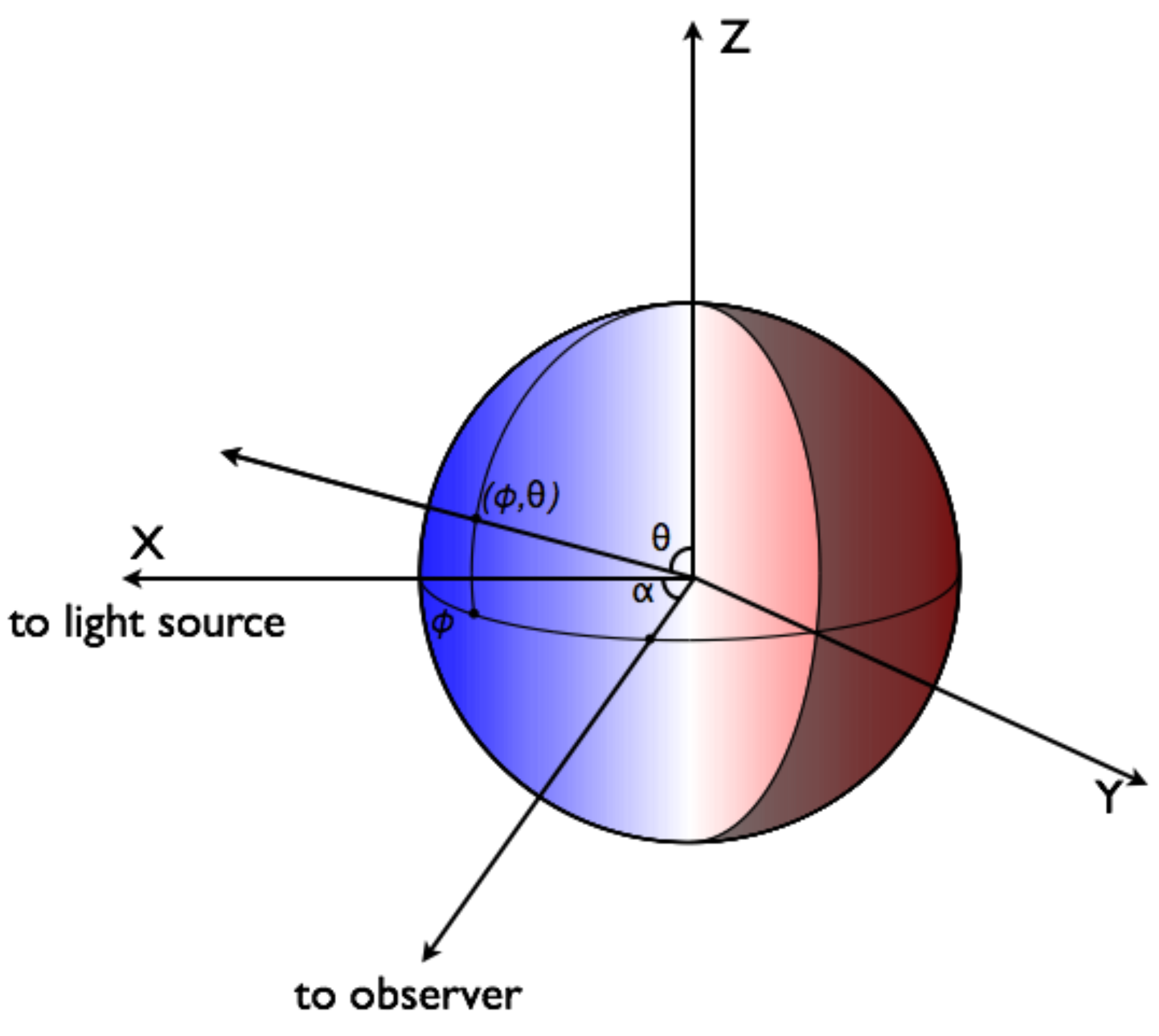} 
   \caption{Geometry used in analytic description of the planetary illumination RV effect. The planet is shown as a sphere being illuminated by a light source along the $X$ axis, which is observed by an observer in the $X-Y$ plane at angle $\alpha$. The planet is rotating with an angular momentum vector in the direction of the $Z$ axis; the color of the planet represents the line-of-sight projected rotational velocity at each point on the surface. The integration is performed over the polar angle $\theta$ and azimuthal angle $\phi$. The setup is adapted from that of \citet{lester}.}
   \label{lestersetup}
\end{figure}

Planets which are detected in reflected light (and are not self-luminous) will show an additional long-period signal as the host star's light illuminates different portions of the planet's surface, with different local rotational velocities, as viewed from Earth. Following \citet{lester}, for a planet which reflects like a Lambertian sphere being illuminated by a uniform light source from the positive $X$ direction, the phase law $\Psi(\alpha)$, or the amount of flux reflected towards an observer at angle $\alpha$ in the $X-Y$ plane, is given by:{\ron \footnote{\ron The following expression is equivalent to Equation 34 of \citet{lester} multiplied by the normalization constant $3/(2\pi)$.}} 

\begin{equation} \label{phasefunctionintegral}
    \Psi(\alpha) = \frac{3}{2\pi}\int_{0}^{\pi} \sin^3{\theta} d\theta \int_{\alpha-\pi/2}^{\pi/2}  \cos{\phi} \cos{(\alpha - \phi )}d\phi 
\end{equation}

\noindent where $\theta$ is the polar spherical angle and $\phi$ is the azimuthal spherical angle, as shown in Figure \ref{lestersetup}. Equation \ref{phasefunctionintegral} is only valid over the interval $0 \leq \alpha \leq \pi$ due to the way the bounds of integration are defined.  To calculate the phase function over the interval $\pi \leq \alpha \leq 2\pi$, the bounds of the $\phi$ integral in Equation \ref{phasefunctionintegral} must be changed so that the integral is evaluated between $\phi = 3\pi/2$ and $\phi = \alpha + \pi/2$. When the integral is evaluated over these limits, the result is: 

\begin{equation} \label{phasefunction}
    \Psi(\alpha) = \sign{(\pi-\alpha)}\frac{\sin(\alpha) + (\pi - \alpha) \cos(\alpha)}{\pi}
\end{equation}

\noindent\ where $\sign{(x)} = 1$ for $x>0$, $\sign{(x)} = -1$ for $x<0$, and $\sign{(x)} = 0$ for $x=0$. Equation \ref{phasefunction} is valid over the interval $0 \leq \alpha \leq 2\pi$. 

\begin{figure*}[t!] 
   \centering
   \includegraphics[width=6.5in]{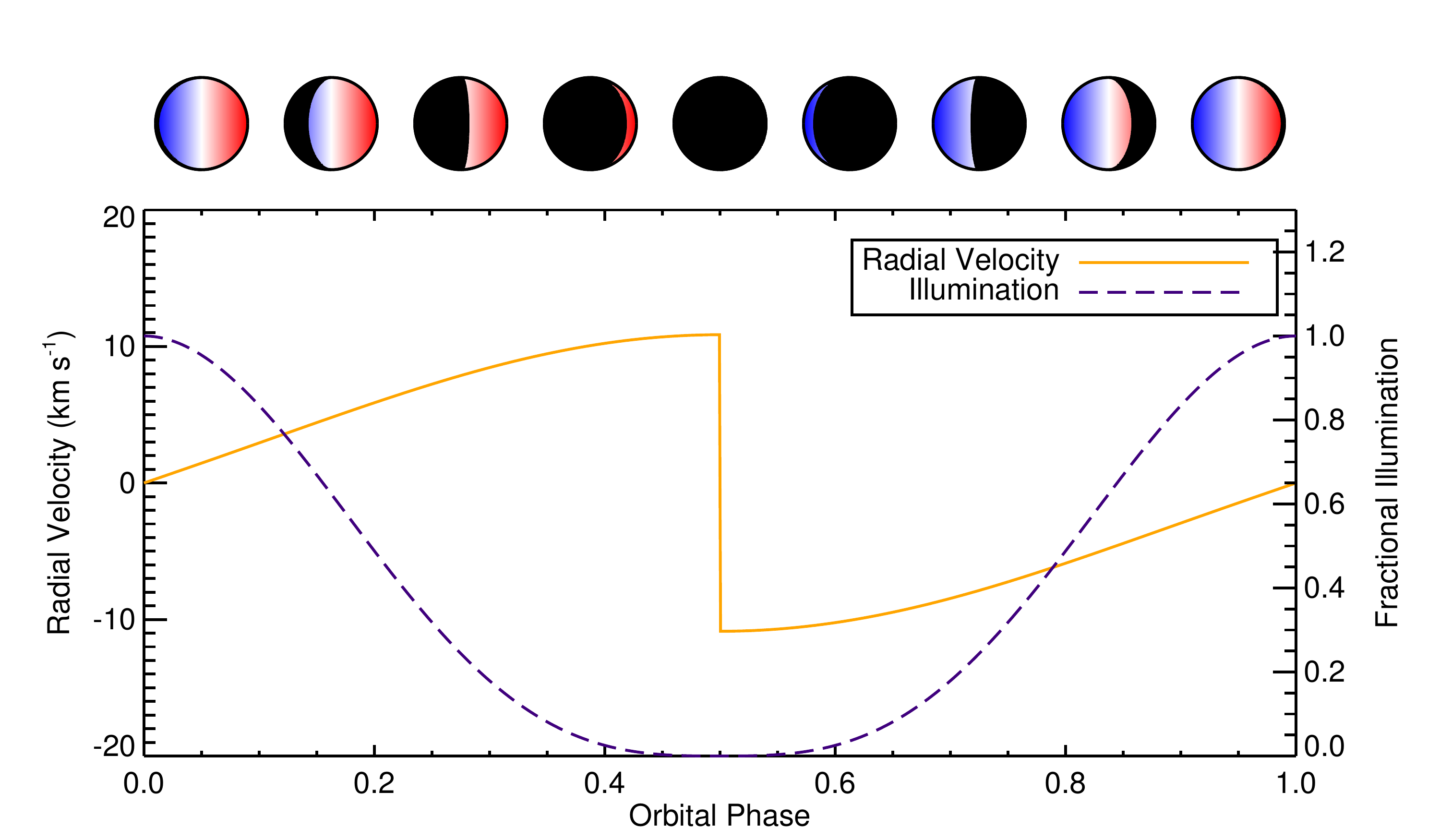} 
   \caption{Brightness and radial velocity signals due to partial illumination of a directly imaged exoplanet detected in reflected light. The radial velocity signal due to the planetary illumination effect as a function of the planet's orbital phase is shown as a solid orange curve, and the brightness of the planet is shown as a dashed purple curve. These curves assume the planet reflects like a Lambertian sphere, the inclination of the planet's orbit is $i_o = 90 ^\circ$, and the planet's spin angular momentum is aligned with the orbital angular momentum (that is, $\beta = \xi = 0$ using the definitions described in the Appendix). Above the curves are diagrams of the illumination of the planet at a handful of orbital phases. The color on the planet's surface shows the radial velocity, where blue is a velocity of $-v_{\rm eq}$, white is a velocity of $0$, and red is a velocity of $+v_{\rm eq}$.}
   \label{illuminationana}
\end{figure*}

The radial velocity signal due to planetary illumination can be calculated by modifying this phase function by including a term describing the radial velocity of the planet's surface, $v_{s}$, at every point $(\theta, \phi)$. We calculate the average $v_{s}$ over the same illuminated region of the sphere: 
%

\begin{equation}
    {\rm RV}(\alpha) = \frac{1}{\Psi(\alpha)} \iint v_{s}\sin^3{(\theta)} \cos{(\phi)} \cos{(\phi - \alpha)}d\phi d\theta
\end{equation}
In the special case where the planet's orbit is viewed edge on ($i_o = 90 ^\circ$) and the rotational axis of the planet is aligned with its orbital angular momentum, $v_{s}$ can be written as: 

\begin{equation}
    v_{s} = -v_{\rm eq} \sin{(\phi - \alpha)} \sin{(\theta)}
\end{equation}

\noindent where $v_{\rm eq}$ is the planet's equatorial rotational speed. Evaluating the integral gives: 
\begin{equation}
    {\rm RV}(\alpha) = v_{\rm eq} \sign{(\pi-\alpha)} \frac{3 \pi \sin{\alpha} (\cos{\alpha} + 1)}{16 (\sin{\alpha} + (\pi - \alpha) \cos{\alpha})}
\end{equation}

\begin{figure*}[t!]
\centering
\includegraphics[width=6.0in]{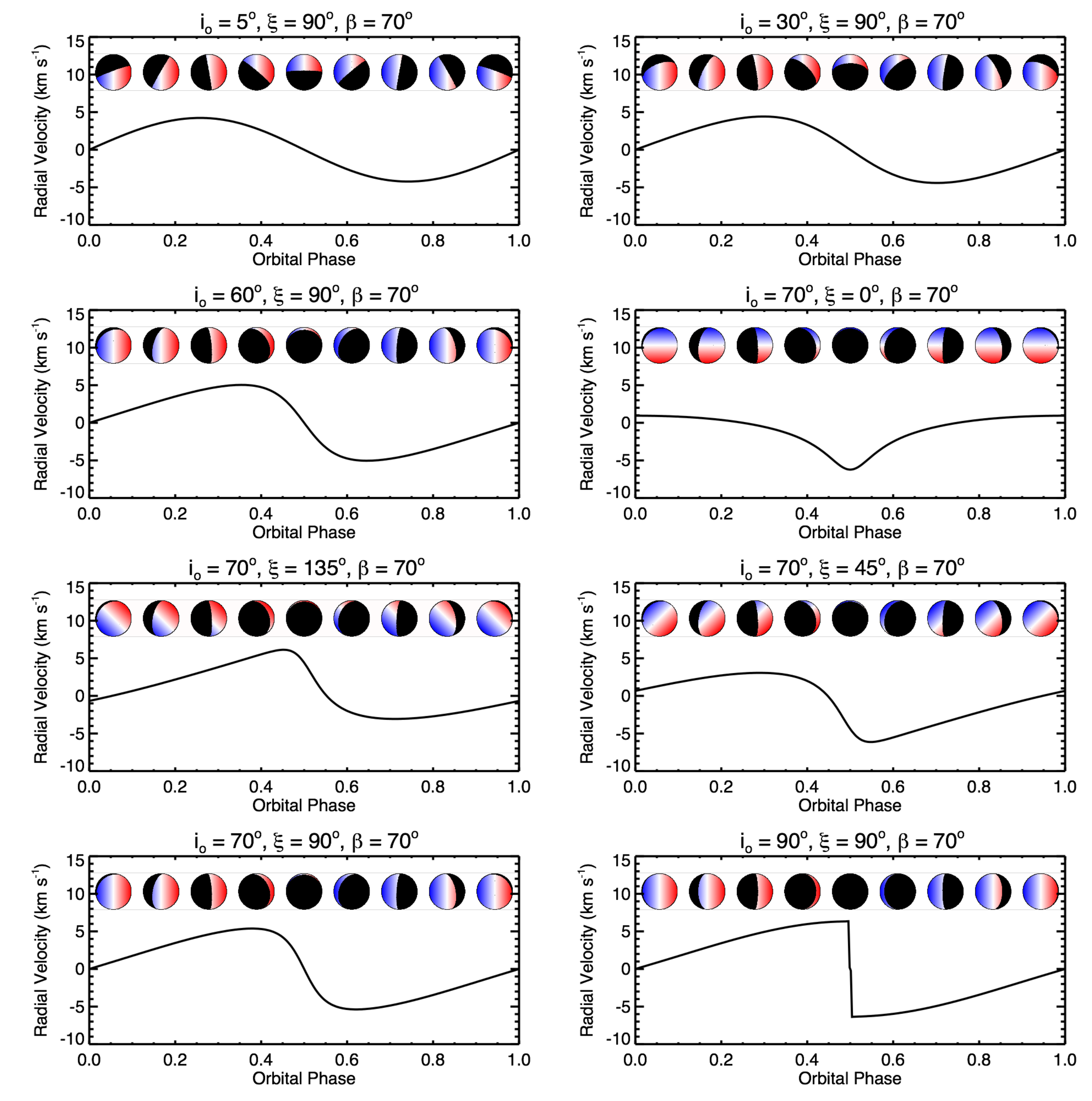}
\caption{Sample planetary illumination RV curves for different spin and orbital orientations. The signal is specified by the planet's orbital inclination $i_o$ and two angles $\xi$ and $\beta$, which describe the orientation of the planet's spin axis. These angles and geometry are described in the Appendix. We have arbitrarily chosen the host planet to be about the size of Jupiter with a rotation period of 18 hours. Above each of the curves are diagrams of the illumination of the planet at a handful of orbital phases. The color on the planet's surface shows the radial velocity, where blue is a velocity of $-v_{\rm eq}$, white is a velocity of $0$, and red is a velocity of $+v_{\rm eq}$.}
\label{generalrv}
\end{figure*}

The illumination and radial velocity curves as a function of phase in this special case are shown in Figure \ref{illuminationana}. In general, when planets are not viewed edge-on ($i_o \neq 90 ^\circ$) and do not have spins which are aligned with their orbits, the illumination effect is more complex and difficult to study analytically. We have developed a framework to calculate the illumination RV effect in general cases numerically, which we describe in more detail in the Appendix. We show a handful of illustrative RV illumination curves in Figure \ref{generalrv}. 

Like the planet's orbit, the planetary illumination RV function is strictly periodic (on the planet's orbital period). The planetary illumination function can have sharp/short timescale features, like the discontinuity shown in Figure \ref{illuminationana}, but these sharp features typically take place when the planet's brightness is very low, making them practically difficult to see. The planetary illumination RV signal should therefore not significantly complicate the detection of exomoons.

\subsection{Planetary Activity Signals} \label{activity}

RV observations of directly imaged exoplanets would also likely show spurious RV variations due to inhomogeneities on the planet rotating in and out of view.  Similar apparent RV variations have been observed and extensively studied on stars \citep{saar, wright, dumusque, boisse, aigrain, rajpaul, haywood}. On stars, these variations are often referred to as ``stellar activity signals'' because the surface inhomogeneities which cause the variations, like starspots, faculae, or plage, are usually the result of magnetic activity.

On the surface of a planet, inhomogeneities like clouds or storms rotating in and out of view on a planet's surface may cause analogous ``planetary activity signals''. We can estimate the impact of planetary activity using similar methods as used to estimate activity signals on stars. The RV variation $\Delta_{\rm RV}$ caused by a dark spot with flux contrast $F_{\rm spot}$ rotating in and out of view on the planet's surface is approximated as: 

\begin{equation}
    \Delta_{\rm RV} \approx F_{\rm spot} \times v\sin{i}
\end{equation}

\noindent where $v\sin{i}$ is the planet's projected rotational velocity. Giant solar system planets \citep{roweneptune}\footnote{The variability can be strongly wavelength-dependent \citep{gelino, stauffer}.} and brown dwarfs \citep[see][and references therein]{wilson, radigan, biller, wow2, vos, artigau} can show variability up to a few percent peak to peak, translating into spot filling fractions of order a few percent. Combining these photometric amplitudes with typical\footnote{Jupiter has a rotational velocity of about 12 \kms, Saturn has a rotational velocity of about 10 \kms, Uranus and Neptune each have rotational velocities of about 2.5 \kms, and $\beta$ Pictoris b has a rotational velocity of about 25 \kms \citep{snellen}.} planetary rotational velocities up to 10-25 \kms\ gives RV variations with amplitudes up to a few hundred meters per second peak to peak. 

\begin{figure*}[ht!] 
   \centering
   \includegraphics[width=6.5in]{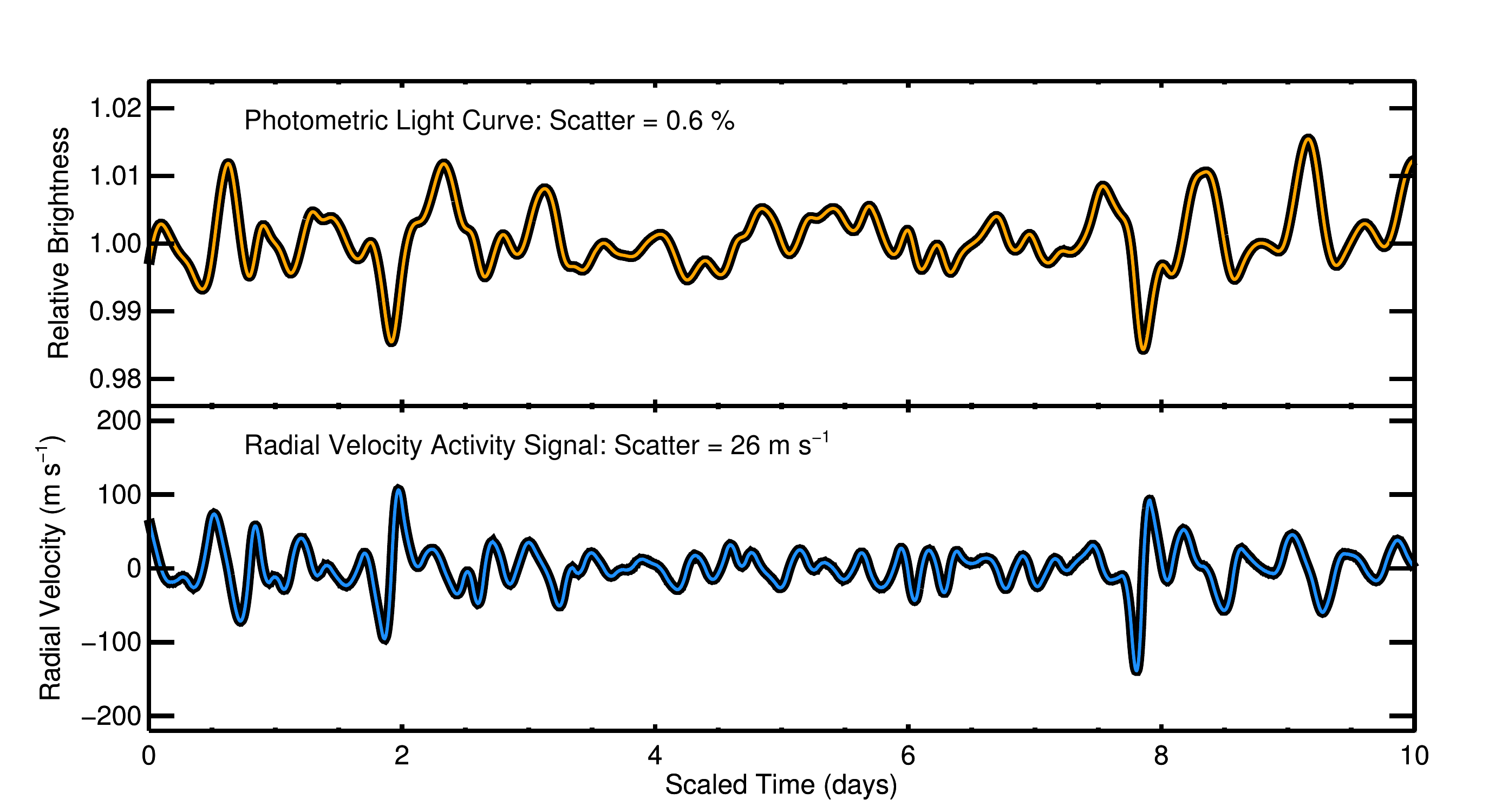} 
   \caption{Simulated RV activity signal for a directly imaged planet. The top panel shows the spline-smoothed brightness of the planet Neptune as a function of time (scaled so that the planet's rotation period is 18 hours), as recorded by the K2 mission \citep{roweneptune}. The bottom panel shows the planetary activity radial velocity signals calculated using the FF' method \citep{aigrain}. The planetary activity signal is calculated assuming the planet is the radius of Jupiter, is self-luminous, and has an 18 hour rotation period. The RMS scatter of the planetary activity signal is about 25 \ms, but larger and more rapidly rotating planets like \betapicb\ can induce higher-amplitude activity signals (see Figure \ref{simulated}).}
   \label{ffprime}
\end{figure*}

Like their stellar counterparts, planetary activity signals present as quasi-periodic radial velocity variability, contributing strong signals at the planet's rotation period and its harmonics. We estimate the form of a planetary activity signal using the FF' method developed by \citet{aigrain}, which is a technique to predict the radial velocity signal caused by dark spots rotating on the surface of a star using photometric observations\footnote{The FF' method also predicts RV variations on stars due to the suppression of convective blueshift in active areas. Since this phenomenon is unlikely to be important on directly imaged planets, we ignore it here.}. Highly precise and continuous photometric observations of planets are rare, but recently, the planet Neptune was observed by the K2 mission for about 50 days \citep{roweneptune}. We fit the K2 light curve of Neptune with a basis spline to smooth out high-frequency noise, and calculate the expected planetary activity signals using FF'. We show the spline-smoothed light curve and resulting planetary activity signal in Figure \ref{ffprime} for the case of a Jupiter-sized planet with Neptune-like photometric variability and a rotation period of 18 hours\footnote{To simulate the activity signals for stars with different rotation periods, we scaled the time axis of the K2 observations by the ratio between the desired rotation period and Neptune's actual rotation period of 16.1 days.}.

The above calculations and discussion are for the case of self-luminous planets. Planets imaged in reflected light will also present quasi-periodic radial velocity variations on the timescale of the planetary rotation period, but because only part of the planetary disk is illuminated, these signals will have somewhat different morphology and lower amplitude.

While planetary activity could produce spurious RV signals with about the same amplitude as the expected RV signal for moons like \thismoon, planetary activity should not prove prohibitive for detecting moons. {\ron In many cases, it is possible to filter and separate activity signals from center-of-mass radial velocity signals when the timescales of these signals are somewhat different \citep[e.g.][]{dumusquegranulation, haywood2014, vanderburg16}\footnote{\ron It is most challenging to separate activity and center-of-mass radial velocity signals when the orbital period is within about 10\% of the characteristic activity timescale or its harmonics (see Figure 4 of \citealt{vanderburg16} and Figure 13 of \citealt{damasso}).}.}  For rapidly rotating planets which will induce the largest spurious RV signals, the rotation periods (typically $\sim$ 10 hours, \citealt{biller, vos, lew, zhou}) should be shorter\footnote{Frequently, the dominant periods of RV activity signals are at the first or second harmonics of the rotation period  \citep{vanderburg16}, further separating the rotation signals from exomoon orbits.} than the orbital periods of most massive exomoons, making it possible to effectively filter away the planetary activity \cite[if RV observations are taken with a high-enough cadence,][]{dumusquegranulation, lopezmorales}. These planetary activity signals should also cause significant changes to the shapes and profiles of spectral lines\footnote{{\ron These line shape diagnostics include the ``bisector span'', a measurement of the skew of the line, the average ``full width at half maximum'' or FWHM width of the spectral features, and others \citep{queloz2001,dumusquefwhm, figueira}. Line shape diagnostics like these sometimes correlate with spurious radial velocity signals caused by activity (see for example Figure 7 of \citealt{queloz2001}). The bisector span is particularly useful for particularly active stars with dark spots, a situation likely analogous to the planetary activity considered here, while the FWHM is more useful for quiet stars like the sun, and may be less sensitive to the large planetary activity signals. Additional diagnostics might be identified, or new techniques might be developed which can translate spectral line shapes into spurious radial velocity signals. See \citet{acc} for a recent review of the subject.}}, which can be used to differentiate activity signals from the signals of exomoons. It should be possible to take advantage of techniques and observing strategies being developed for mitigating stellar activity for RV planet searches and apply them to RV moons searches \citep[e.g.][]{haywood2014, rajpaul, donati}.

\subsection{Peak-pulling by Light from the Exomoon} \label{peakpulling}

While detecting exomoons by Doppler monitoring of directly imaged planets is in many ways similar to detecting exoplanets by Doppler monitoring of stars, the analogy is not perfect. One difference between detecting exoplanets and exomoons with radial velocities is that exomoons may be bright enough compared to their host planets to contribute significant light to the planetary spectrum, which can perturb the measured radial velocity of the planet. This effect, which is known as peak-pulling, has been studied in the case of double-lined spectroscopic binary stars and has been shown to significantly affect radial velocity measurements if not taken into account \citep{todcor}. Because exoplanets are at least thousands (and often millions) of times fainter than their host stars, this effect can generally be ignored for exoplanet detection. 

In the case of a planet/moon system imaged in reflected light, we can write the brightness ratio $\rho_b$ between the planet and moon as:

\begin{equation}
    \rho_b = \frac{\alpha_p r_p^2}{\alpha_{\leftmoon} r_{\leftmoon}^2}
\end{equation}

\noindent where $r_p$ and $\alpha_p$ are the radius and albedo of the planet, and $r_{\leftmoon}$ and $\alpha_{\leftmoon}$ are the radius and albedo of the moon. Assuming similar albedos, a Jupiter-sized planet imaged in reflected light will therefore only be about 8 times brighter than its Neptune-sized exomoon, close enough in brightness that the moon might significantly affect the measured planetary radial velocity. If peak-pulling is not accounted for in the radial velocity extraction, the measured radial velocity (${\rm RV}_{\rm measured}$) will be the brightness-weighted {\ron average} of the radial velocity of the planet (${\rm RV}_p$) and moon (${\rm RV}_{\leftmoon}$):

\begin{equation}
    {\rm RV}_{\rm measured} = \frac{{\rm RV}_{p} \rho_b + {\rm RV}_{\leftmoon}}{1 + \rho_b}
\end{equation}

The radial velocity of the moon about the center of mass of the planet/moon system is a Keplerian function with the same orbital parameters but opposite direction as the planet's motion, with a semiamplitude:

\begin{equation}
    K_{\leftmoon} = K_p \frac{m_p}{m_{\leftmoon}}
\end{equation}

If the contribution from the moon is not taken into account, the measured radial velocity will therefore be a Keplerian function with a semi-amplitude $K_{\rm measured}$ given by:

\begin{equation}
    K_{\rm measured} = K_p \frac{\rho_b - m_p/m_{\leftmoon}}{1 + \rho_b}
\end{equation}

In many cases, including the case of a Neptune-like moon orbiting a Jupiter-like planet detected in reflected light, the radial velocity contribution from the moon's light might actually dominate over the signal from the planet's light. If measured without taking the moon's contribution into account, the radial velocity signal detected will be opposite in sign and significantly different in amplitude from the motion of the planet.

In practice, peak-pulling will only take place when the radial velocity of the moon is close enough to the radial velocity of the planet that spectral features from the planet and moon overlap in the combined spectrum. In particular, peak-pulling will only be important for the parts of the planet/moon orbit near conjunctions, when the radial velocity of the moon is within about $v\sin{i}$ of the radial velocity of the planet. Even when spectral features from the planet and moon do overlap, analysis techniques have been developed over the years \citep[e.g. ][]{todcor, czekala} to disentangle the contributions from each component of a double-lined spectrum. If these spectral features can indeed be disentangled, then the secondary lines from the moon could yield the true mass ratio between the planet and moon, helping to clarify the moon's nature. 

Peak pulling is likely to be less important for self-luminous planets, since massive planets should cool more slowly (and therefore stay luminous longer) than lower-mass objects. However, one could envision scenarios where the moon formed after the planet, making it possible to detect spectral features from a self-luminous moon. 

\subsection{Partial Occultation by Disk Clumps} \label{diskclumps}

Some young, self-luminous planets may still be embedded within the protoplanetary disks from which they formed. Observations of young stars have shown that in many cases, protoplanetary disks have optically thick clumps which can occult the parent stars, quasi-periodically blocking significant fractions of the star's light \citep[e.g.][]{cody, rodriguez}. These dips can repeat on timescales from days to decades, and can cause eclipses which last for timescales ranging from hours to years \citep{rodriguez2}. Disk clumps that occult a rotating planet might cause radial velocity signals by introducing asymmetries in the planet's rotation profile.

To estimate the radial velocity signal caused by disk clumps occulting a young, self-luminous planet, we assume that clumps in the disk have roughly the same properties close to the host star and far away where directly imaged planets might orbit. \citet{cody} reported observations of ``dipper'' stars, which undergo roughly day-long occultations by disk clumps which repeat quasi-periodically on timescales of order 5-10 days. The Keplerian\footnote{The dust clumps may not be strictly undergoing Keplerian motion if they are trapped by the star's magnetic field near the corotation radius, but even in this case the clump velocity will still be close to Keplerian.} velocity of a disk clump, $v_{\rm clump}$, which orbits with a quasi-period, $P_q$, is given by: 

\begin{equation}
    v_{\rm clump} = \left [ \frac{2 \pi G M_\star}{P_q} \right]^{1/3}
\end{equation}

For a typical dipper, which has a quasi-periodicity of about a week and orbits a star with $M_\star \approx$ 0.5 \msun, the Keplerian velocity of the clump is about 80 \kms. Combined with a dip duration, $t_{\rm dip}$, of about a day, this velocity implies the clump size,

\begin{equation}
r_{\rm clump} \sim \frac{v_{\rm clump} t_{\rm dip}}{2}
\end{equation}

\noindent is likely of order 5 \rsun, considerably larger than the host stars. Since the dips only block 30-50\% of the star's light, the clumps are likely not optically thick. 

If similar disk clumps exist farther away from the host star, at orbital distances of 1-50 AU around more massive stars like those which have been found to host directly imaged exoplanets, the Keplerian velocities will be smaller, on the order of 5-30 \kms. Occultations of stars (and directly imaged planets orbiting far out in the disk) by these long-period disk clumps would last for a few days to a few weeks. 

Any velocity signal induced by an occulting disk clump would be due to an asymmetry in the flux received from the approaching hemisphere and the receding hemisphere of the rotating planet. We estimate this asymmetry by calculating the change in the flux transmitted through the clump over a distance from the center of the clump, $\frac{dI}{dx}$. We write: 

\begin{equation}
    \frac{dI}{dx} \sim \frac{dI}{dt}\frac{dt}{dx} \sim \frac{dI}{dt} \frac{1}{v_{\rm clump}}
\end{equation}

Here, $\frac{dI}{dt}$ is just the time derivative light curve of a dipper star. For typical dippers in short-period orbits with $v_{\rm clump} \sim 80$ \kms and which cause 30-50\% drops in flux on timescales of 12 hours to a day, $\frac{dI}{dt}$ is a few percent per hour. Combining this value with the orbital velocity gives $\frac{dI}{dx} \sim 10^{-5}$\% per km, or about 1\% per Jupiter radius. 

The velocity signal caused by this flux gradient due to an occulting disk clump is therefore given by: 

\begin{equation}
    {\rm RV_{\rm clump}} \sim   r_p \frac{dI}{dx} v\sin{i}
\end{equation}

\noindent where $r_p$ is the radius of the planet. For a Jupiter-sized planet with an 18 hour rotation period, the RV signal from disk clump occultations will have an amplitude somewhere around 10-50 \ms, while a larger, more rapidly rotating planet like \betapicb\ could show signals up to a few hundred meters per second. The timescales for these signals would be days to weeks (the crossing time for the clumps over the planet assuming relative velocities close to the Keplerian orbital velocity), and the disk occultation signals would likely not repeat periodically, since the occulting clumps show evolution on timescales much shorter than the orbit of a directly imaged planet. 

Disk clump occultations will only be a concern for the very youngest directly imaged planets, whose host stars still retain their protoplanetary disks. Even the young exoplanet \betapicb\ is too old for disk clump occultations to be present; the star, \betapic, only hosts a debris disk. Even around very young stars where disk clump occultations could introduce stochastic variability into planetary radial velocity time series, the signals introduced should not preclude the detection of massive exomoons. Like ``planetary activity'' signals, disk clump occultations do not produce RV signals with amplitudes greatly exceeding the amplitudes of RV signals from massive exomoons. Disk clump occultations should also be identifiable by changes to the planetary line profile. Finally, statistical techniques designed to aid in the detection of low-mass planets around stars should help to separate out the stochastic signals from disk clump occultations from the periodic signals caused by exomoons. 

\section{Detection Feasibility}\label{detectability}

In this section, we estimate the feasibility of detecting exomoons with Doppler monitoring. We base our estimates on the successful detection of thermal light from $\beta$ Pictoris b by \citet{snellen}. \citet{snellen} observed $\beta$ Pictoris b with the Cryogenic High-Resolution Infrared Echelle Spectrograph (CRIRES) behind the Multi-Application Curvature Adaptive Optics (MACAO) system on the Very Large Telescope (VLT). At the time of the observations,  $\beta$ Pictoris b was located about 0.4 arcseconds from its host star, and the adaptive optics system attenuated the light from $\beta$ Pictoris by factors between 8 and 30 at the position of the planet. \citet{snellen} detected spectral features from $\beta$ Pictoris b with a signal-to-noise ratio of 6.4 after a total exposure time of about 30 minutes\footnote{The total observing time was closer to an hour after including overheads \citep{snellen}.}. 

From their low signal-to-noise detection of spectral features from $\beta$ Pictoris b, \citet{snellen} were able to measure the radial velocity of $\beta$ Pictoris b with a precision\footnote{\citet{snellen} report an uncertainty of 1.7 \kms\ on the RV of $\beta$ Pictoris b, but included a 0.7 \kms\ systematic uncertainty term based on the uncertainty of the absolute RV, so the photon-limited Doppler precision is closer to 1.5 \kms.} of about 1.5 \kms. 

Detecting exomoons inducing RV semiamplitudes of about 200 \ms\ is not feasible when the precision of an individual measurement is limited to 1.5 \kms, but fortunately, higher precision RV measurements are possible. {\ron Following \citet{lovisfischer}, the photon-limited uncertainty of a radial velocity observation, $\sigma_{\rm RV}$, scales as:}

\begin{equation}
    \sigma_{\rm RV} \propto \frac{\rm FWHM^{\ron 3/2}}{S/N}
\end{equation}
\noindent where ${\rm FWHM}$ is the width of the lines as seen by the spectrograph\footnote{This term includes both the natural broadness of the spectrum and instrumental line broadening.} and $S/N$ is the signal-to-noise ratio of the detection. In particular, $S/N$ is proportional to: 

\begin{equation}
    S/N \propto \sqrt{S} \sqrt{N_{\rm lines}}
\end{equation}

\begin{figure*}[ht!] 
   \centering
   \includegraphics[width=6.5in]{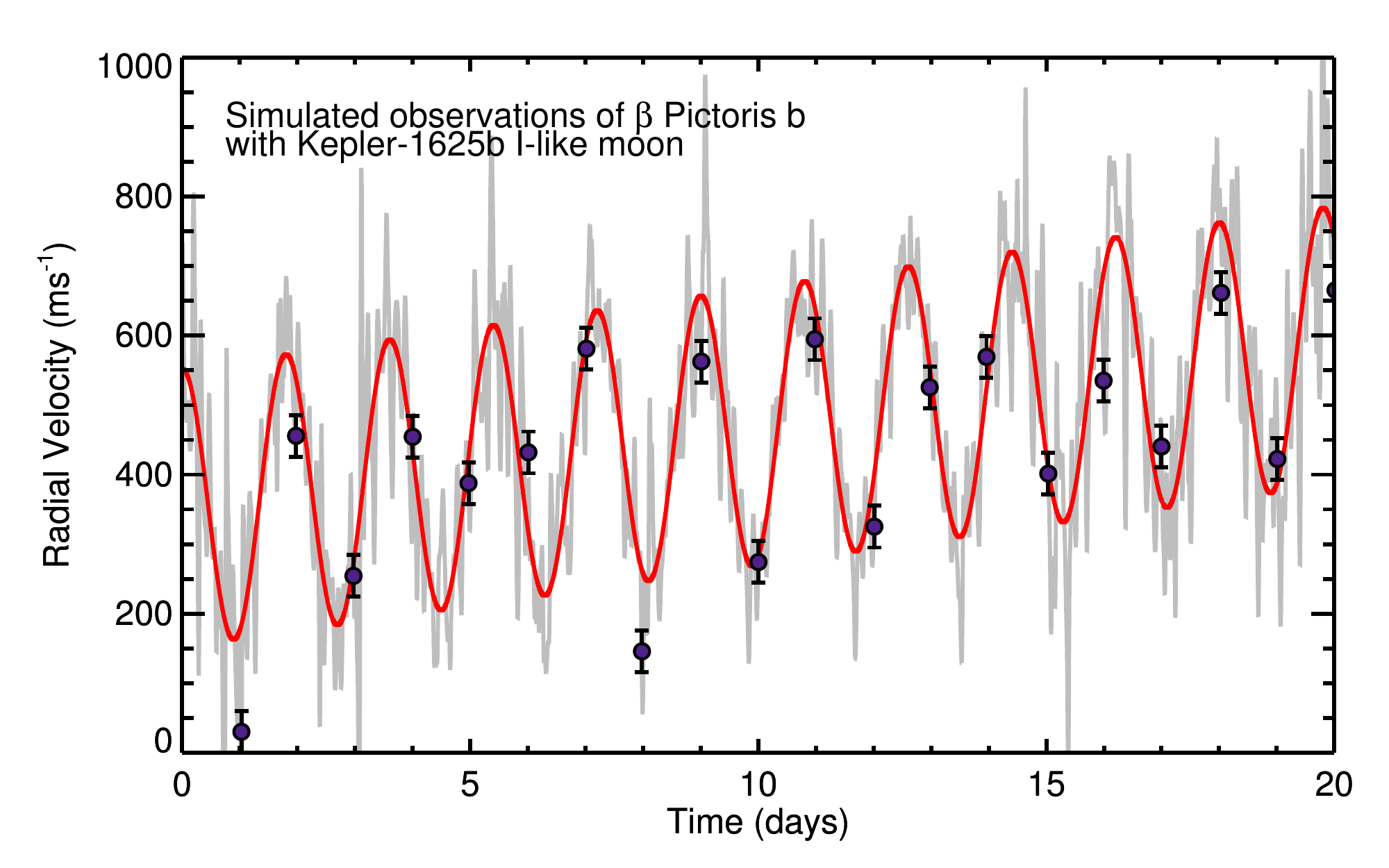} 
   \caption{Simulated observations of the self-luminous directly imaged planet \betapicb, assuming it hosts a \thismoon-like exomoon. The purple points simulate a nightly observing cadence, and assume 30 \ms\ RV photon-limited uncertainties (see Section \ref{detectability}). The red curve is the center-of-mass motion of the planet; the long-term trend is due to the orbit of the planet about its host star, and the fast sinusoid is due to the orbit of the moon about the planet. The grey curve is the total expected RV curve, including ``planetary activity'' caused by inhomogeneities rotating in and out of view on the surface of the planet. We calculated the planetary activity signal using the FF' method described by \citet{aigrain}, assuming a planetary radius of 1.45 \rj\ \citep{morzinski}, and photometric variability from K2 obserations of Neptune \citep{roweneptune}, with the timescale of the variations scaled from Neptune's 16 hour rotation period to \betapicb's likely 7 hour rotation period. The RV signal due to the exomoon in this case is clearly detected {\ron (with a bootstrapped false alarm probability of $p\approx5\times10^{-4}$)} in only three weeks of nightly observations.}
   \label{simulated}
\end{figure*}

\noindent where $S$ is the suppression factor of starlight at the position of the planet compared to what would be observed at the center of the PSF and $N_{\rm lines}$ is the number of spectral lines observed\footnote{\ron The number of lines is typically in the range of $10^3-10^4$ for broadband spectrographs observing stars. Both self-luminous and reflected-light planets will likely have more lines due to the presence of molecular absorption in the spectra.}. A natural way to improve the RV precision in a given observation is to increase the signal-to-noise of the observations\footnote{Alternatively, the precision of RV observations could be improved by observing a planet that rotates more slowly than $\beta$ Pictoris b and therefore has narrower spectral features. Indeed, observations of GQ Lupi b \citep{schwarz}, which is more slowly rotating than \betapicb, achieved RV precision of about 400 \ms.}, which should be possible with modern instrumentation capabilities.

One way to improve the detection strength would be to increase the bandpass of the instrument used to observe radial velocities so that it can observe more planetary spectral features. \citet{snellen} obseved with CRIRES, which only had one spectral order (which for these observations was centered on the CO bandhead at 2.3 microns). The upgraded version of CRIRES, CRIRES+ \citep{criresplus}, will be able to cover 10 times the bandpass as the original instrument, significantly increasing the number of planetary spectral features to use in an RV analysis. If all spectral orders are equal in terms of the number and depth of lines, a boost of a factor of 10 in spectral bandpass could improve the S/N by a factor of roughly 3; realistically, not all spectral orders will have as many deep and sharp lines as the CO bandhead, so the improvement will not be as great. 

Another way to increase the strength of the detection would be to use modern high-contrast adaptive optics imaging and coronagraphy to separate the starlight from the light of the planet. The observations conducted by \citet{snellen} achieved a starlight suppression of about a factor of 8-30 at a distance of 0\farcs4 from the position of $\beta$ Pictoris b. Modern adaptive optics systems equipped with coronagraphs, like GPI \citep{gpi} and SPHERE \citep{sphere}, can significantly improve upon this level of starlight suppression. According to the SPHERE instrument handbook\footnote{See Figure 13 of \url{https://www.eso.org/sci/facilities/paranal/instruments/sphere/doc/VLT-MAN-SPH-14690-0430_v95.pdf}}, on a bright star in imaging mode, SPHERE can suppress starlight by factors of a few thousand or more at distances of 0\farcs4, and can yield even greater suppression (factors of up to $10^4$) when a coronagraph is used. Compared to the \citet{snellen} observations, using a modern high contrast AO system could yield a signal-to-noise boost of a factor of 10, and potentially up to a factor of 30 in S/N. 

Altogether, a one-hour observation similar to that conducted by \citet{snellen} using a CRIRES+-like spectrograph behind a SPHERE-like adaptive optics system could yield a detection of \betapicb\ with signal-to-noise somewhere between 120 and 300, which would yield a photon-limited Doppler precision on \betapicb\ between 30 \ms\ and 75 \ms, sufficient to detect massive exomoons like \thismoon. We show a simulated RV detection of a \thismoon-like moon orbiting \betapicb\ in Figure \ref{simulated}. 

While it should be possible to detect planetary spectral features with high enough signal-to-noise to measure precise radial velocities on planets like \betapicb\ with existing telescopes and modern instrumentation, we note that this capability should be dramatically increased when the next generation of 30-meter-class telescopes come on line.  All else being equal, the signal-to-noise ratio of a direct imaging detection of a planet is proportional to the diameter of the telescope squared. Compared to the observations performed by \citet{snellen} on the 8 meter VLT, a similar observation a 30-meter-class telescope (such as using GMTNIRS on the Giant Magellan Telescope, \citealt{jaffe})  would yield a signal-to-noise roughly $(30/8)^2 \approx 15$ times higher. Additionally, starlight suppression on 30 meter class telescopes should be much better than on 8 meter class telescopes at a given separation on the sky, because the separation will be greater in terms of the telescope's resolution elements. In fact, \citet{snellen} argue that thanks to the higher resolution of 30 meter class telescopes, scattered light from $\beta$ Pictoris should have negligible impact on spectroscopic observations of \betapicb, making it possible to obtain high-quality planetary spectra with only short observations. When combined with the expected improvement in the performance of high-contrast imaging systems in the coming years, 30-meter-class telescopes should make Doppler surveys of large numbers of directly imaged planets feasible \citep{eltao}.

{\ron Detecting exomoons around planets imaged in reflected light will be more difficult. So far, no exoplanets have been detected in reflected light by direct imaging\footnote{\ron Reflected light from short-period exoplanets has been detected by \Kepler\ and CoRoT \citep[e.g.][]{alonso, boruckihatp7, sanchisojedak78, malavolta}.}, but improvements to adaptive optics systems and the construction of 30 meter class telescopes are expected to lead to detections of these planets (\citealt{naswhitepaperarxiv}\footnote{\url{http://surveygizmoresponseuploads.s3.amazonaws.com/fileuploads/15647/4139225/235-ba43dcc83ed8463dd0e7af6cb4f510ba_FitzgeraldMichaelP.pdf}}, J. Males et al. \textit{in prep}). These planets detected in reflected light are likely to be quite faint, with nearby ($\approx$ 10 pc) giant planets in Jupiter-like orbits having optical apparent magnitudes around 25, which will make precise RVs difficult but not impossible. Using the 10m Keck I telescope, \citet{jj} measured radial velocities of a V=17 star with precision of 20 \ms\ in 20 minute exposures. Scaling these observations to a 30m class telescope, with a wider spectral bandpass (\citealt{jj} used an iodine cell to calibrate their radial velocity observations which limited their bandpass to about 100 nm) and higher spectral resolution expected for future generations of precise spectrographs, it should be possible to measure RVs with a bit lower precision ($\approx 100$ \ms) on 26th magnitude stars in one hour exposures. Also, planets more readily detectable in reflected light than Jupiter may be discovered -- the brightness of a planet in reflected light is inversely proportional to its orbital distance squared, so closer planets to their host stars will be much easier to detect \citep{traub}. However, close-in exoplanets may be less likely to host exomoons than more distant exoplanets because requirements for dynamical stability closer to the star are more stringent.}

\section{Discussion} \label{discussion}

So far, we have shown that massive exomoons like the candidate moon around \thisplanet\ could induce large RV variations on the moon's host planet, and we have shown that RV variations of this amplitude could be detected with present-day or forthcoming instrumentation. Here, we argue that a radial velocity survey of directly imaged exoplanets to detect massive exomoons is a worthwhile endeavor, and we describe additional science which might come from such a survey. 

First and foremost, if an RV exomoon survey of directly imaged exoplanets is successful, the  detection of an exomoon (or exomoons) would provide new and important knowledge about planetary systems that could not be accessed in any other way. For example, either the discovery of massive exomoons, or limits placed on the presence of such moons, would directly inform models of planet and moon formation. It is generally believed that moons as large as the proposed \thismoon\ cannot form via in-situ accretion in a circumplanetary disk; the mass ratio of \thismoon\ to its host planet is too large to be formed in this way. Therefore, the presence of moons like \thismoon\ around giant planets would likely indicate that some dynamic process (either a capture process or a giant impact) took place. If a population of massive \thismoon-like exomoons were to be found, it would, like the discovery of hot Jupiters, indicate that planet (and moon) formation is a more dramatic and eventful process than previously thought. The successful detection of massive exomoons could turn planetary formation theory on its head the way the detection of hot Jupiters did decades ago. 

Even if massive exomoons turn out to be rare or don't exist, there are strong scientific motivations to conduct an RV survey of directly imaged planets. One such motivation would be detecting planetary activity signals due to the surface inhomogeneities rotating in and out of view on the planets' surfaces. While these signals are a nuisance to detecting center-of-mass radial velocity variations like those caused by exomoons, the planetary activity signals encode the rotation period of the planet. When combined with a spectroscopic estimate of the projected rotational velocity (from the broadening of spectral features) and planet radii, these measured planetary rotation periods could yield some of the first measurements of line-of-sight exoplanet obliquities (angles between the planet orbits and spin axes). A survey which yields the rotation periods for a sample of directly imaged planets could begin to assess whether strong dynamical interactions like those which tilted Uranus in our own Solar system are common or rare. 

Additionally, measurements of the obliquities of planets detected in reflected light might be obtained by measuring the planetary illumination RV effect, which encodes the planet's obliquity both in the line-of-sight and sky-projected directions. The planetary illumination RV effect could yield even more information than combining rotational velocities and rotation periods if it is possible to disentangle the signal from the planet's orbital motion. One way to disentangle the signals is to take advantage of the fact that the planetary illumination effect is only present in reflected-light observations. Comparing RV observations of the same planet in the visible, where reflected light is dominant, and far enough in the infrared that even cool planets are self-luminous could cleanly separate these two signals, making obliquity measurements possible.

Another outcome from an RV survey of directly imaged planets would be to very precisely trace out the spectroscopic orbit of self-luminous planets (and planets imaged in reflected light if the orbital signal can be disentangled from the planetary illumination effect). The RV signals due to the planets' orbits around their host stars have amplitudes much higher than the RV precision necessary for the survey, so these observations could yield highly precise orbital elements. For directly imaged planets around stars for which precise radial velocities are difficult or infeasible (like rapidly rotating and active young stars), measuring the planet's spectroscopic orbit could refine orbital elements like period and eccentricity, and could yield a precise dynamical mass measurement for the host star. Measuring the spectroscopic orbit of a planet which has already been detected in radial velocity monitoring of its host star could yield model independent masses for both objects.

Finally, an important byproduct of an RV survey of directly imaged planets will be very high signal-to-noise co-added planetary spectra. We estimate that with present-day instrumentation, it should be possible to detect spectral features in \betapicb\ at a significance of 120-300 $\sigma$ in an hour of observations. Over the course of a Doppler survey, there may be 50-100 individual observations of the planet with this quality. Co-adding all of these spectra could yield some of the highest-quality spectra ever taken of exoplanets. High quality spectra like these could lead to the detection of trace elements in the planet atmospheres.

\section{Summary} \label{conclusions}

In this paper, we investigate the feasibility of detecting exomoons by conducting a Doppler survey of directly imaged exoplanets. We drew inspiration from the detection of a massive candidate exomoon (about the size and mass of Neptune) around the planet \thisplanet. If the candidate exomoon around \thisplanet\ is confirmed, it would be the first in a new class of moons unlike anything seen before in our Solar System. 

We suggest that in analogy to the discovery of the first exoplanets orbiting sun-like stars \citep{struve, mayor}, some of the first exo-moons discovered, like the proposed \thismoon, might be massive and orbiting in short periods around their host planets, making them well suited for detection via radial velocity monitoring. We estimate the amplitudes and timescales of radial velocity signals that might be present in observations of directly imaged planets, and find that the RV semiamplitudes induced by massive moons like \thismoon\ could range from a few hundred meters per second up to a kilometer per second. These signals from massive exomoons are large enough that astrophysical nuisance signals and instrumental stability should not prevent their detection. 

Based on previous spectroscopic observations of directly imaged exoplanets, we estimate that it should be possible to measure radial velocities with fairly high photon-limited precision with reasonable exposure times on bright directly imaged exoplanets. We estimate that using a wide-bandpass high-resolution near infrared spectrograph like CRIRES+ or IGRINS on an 8 meter class telescope behind a modern high-contrast adaptive optics imaging system like GPI or SPHERE, it should be possible to attain photon-limited Doppler precision between 30 and 75 \ms\ on \betapicb, sufficient to detect the RV signal caused by a \thismoon-like moon. The observations required to detect giant exomoons around directly imaged planets should yield additional scientific byproducts, including high signal-to-noise planet spectra and measurements of the spin axis orientation (obliquity) of these directly imaged exoplanets. 
\\

{\ron \noindent \textbf{Note in Review:} Since this manuscript was completed, we have become aware of some impressive analytic work by \citet{kawahara} on the planetary illumination/spin effect.  He was able to carry out the illumination-phase spin radial-velocity integral (RV$(\alpha)$) analytically for arbitrary orbital inclinations as well as planetary spin vectors for a couple of simple scattering laws.  We have verified that his analytic expression for the case of Lambertian scattering does indeed match our numerical integrations.  However, for more complicated and realistic planetary atmospheric scattering functions, the numerical approach we describe is likely to be necessary.}

\acknowledgments
We thank Brendan Bowler, Cyndi Froning, Eric Gaidos, Adam Kraus, Caroline Morley, Sam Quinn, Aaron Rizzuto, and Joey Rodriguez for valuable conversations. We are indebted the anonymous referee for a constructive report which improved this paper. This work was performed in part under contract with the California Institute of Technology (Caltech)/Jet Propulsion Laboratory (JPL) funded by NASA through the Sagan Fellowship Program executed by the NASA Exoplanet Science Institute. This paper includes data collected by the \Kepler\ mission. Funding for the \Kepler\ mission is provided by the NASA Science Mission directorate.

Facilities: \facility{ADS, CDS}

\appendix \label{appendix}
\renewcommand{\theequation}{A.\arabic{equation}}
\renewcommand\thefigure{A.\arabic{figure}}    
\setcounter{figure}{0}  

In this appendix, we describe a numerical scheme to determine the integrated radial velocity of the illuminated surface of a planet imaged in reflected light. We perform the integration numerically by randomly sampling points on the illuminated hemisphere of an exoplanet, calculating the velocity contribution from each point on the hemisphere, and summing the contributions. For this process, we need to know (1) the radial velocity of each point on the illuminated surface of the planet, (2) the brightness of that point for different assumed scattering laws, and (3) whether, in fact, that point is visible to a distant observer.  The velocity of an arbitrary point on the surface of the planet can be written schematically as: 

\begin{equation} \label{appendixtest}
\vec{V} = \vec{V}_o + \vec{V}_{p} +\vec{V}_s
\end{equation}
where $\vec{V}_o$ is the orbital velocity of the center of mass (`CM') of the planet/moon (or binary planet) system in its `outer' orbit about the host star; $\vec{V}_{p}$ the orbital velocity of the primary planet about the CM of the planet/moon system (which we assume here to be coplanar with the outer orbit); and $\vec{V}_s$ the velocity associated with the inertial spin rate of the primary planet about its center. 

\begin{figure*}[t]
\centering
\includegraphics[width=0.7\columnwidth]{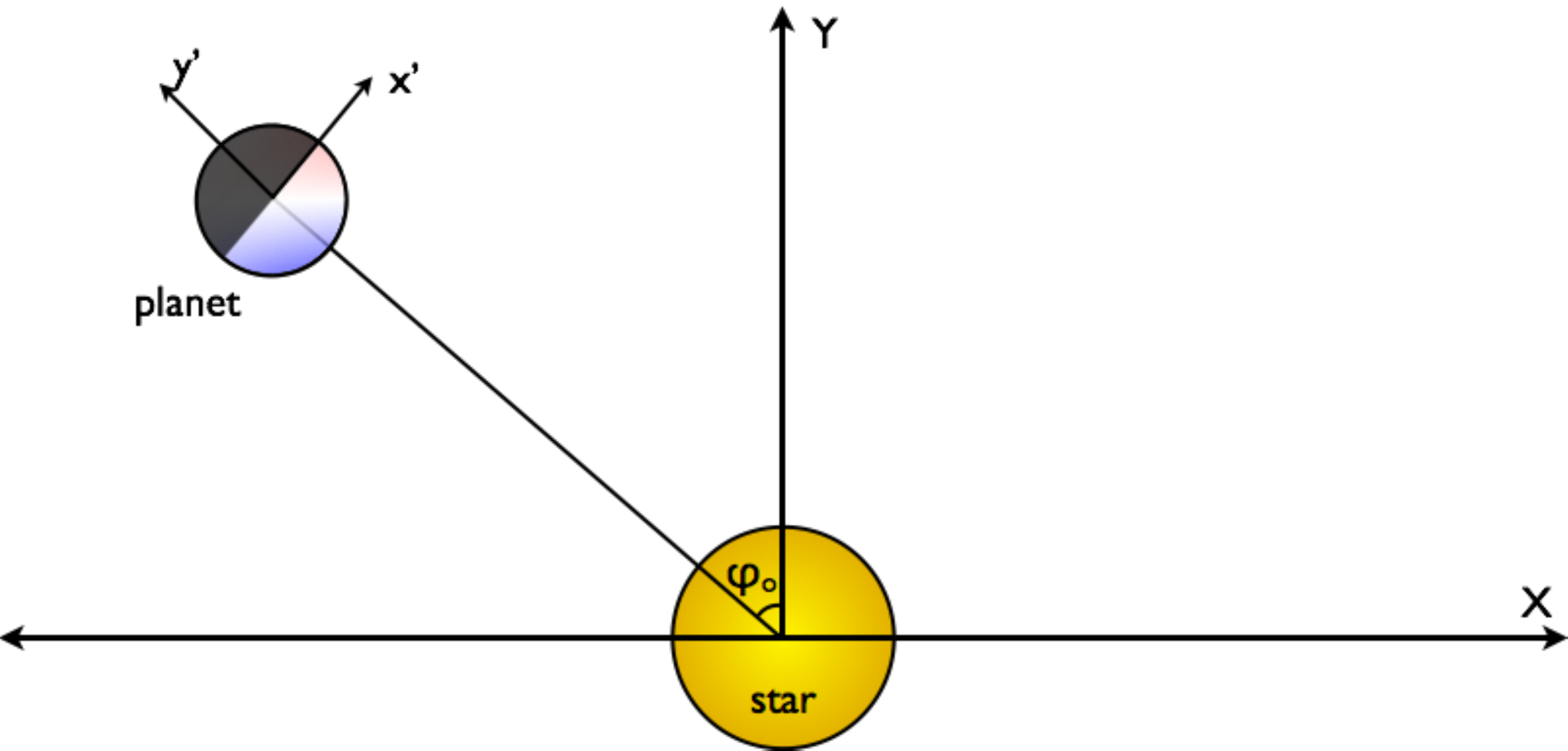}
\caption{Schematic of the illuminated planet geometry, not to scale.  The host star is located at the origin of the $\{X,Y\}$ coordinate system.  The `illumination coordinates', along with the true anomaly of the outer orbit relative to the eclipse, $\varphi_o$, and the outer orbital inclination angle, $i$, determine the planetary phase. The planetary hemisphere facing the host star contains the set of normal vectors $\hat{n}$, whose origin is at the planet center and whose constant vector components are with respect to the $\{x',y',z'\}$ coordinate frame. The two frames are both right-handed, so the $Z$ and $z'$ axes come directly out of the page/screen.}
\label{fig:fig1}
\end{figure*}

{ The velocity of the CM of the planet/moon system is given by
\begin{equation}
\vec{V}_o = -\frac{\Omega_o a_o}{\sqrt{1-e_o^2}}\left[\left\{ \cos (\varphi-\varphi_e)+e_o \sin \omega \right\} \hat{X} +\left\{ \sin(\varphi-\varphi_e) -e_o \cos \omega_o\right\}  \hat{Y} \right] \,
\end{equation}
where $\Omega_o a_o/\sqrt{1-e_o^2}$, $e_o$, and $\omega_o$ are the conventional radial velocity semiamplitude ($K_o$, see Equation \ref{outerorbitsemiamplitude} in the main paper), eccentricity, and argument of periastron, while $\varphi$ is the true anomaly of the planet/moon CM in its outer orbit, and $\varphi_e$ is the true anomaly at the time of superior conjunction (or eclipse).  The geometry of the $X-Y$ plane is specified in Figure~\ref{fig:fig1}.  This form for $\vec{V}_o$ is especially useful in the case of orbits with small eccentricity.  If, on the other hand, we wish to deal with substantial eccentricities, then we can simply revert to the more usual form by setting $\varphi_e+\omega = \pi/2$.}

\begin{figure*}[t]
\centering
\includegraphics[width=0.95\columnwidth]{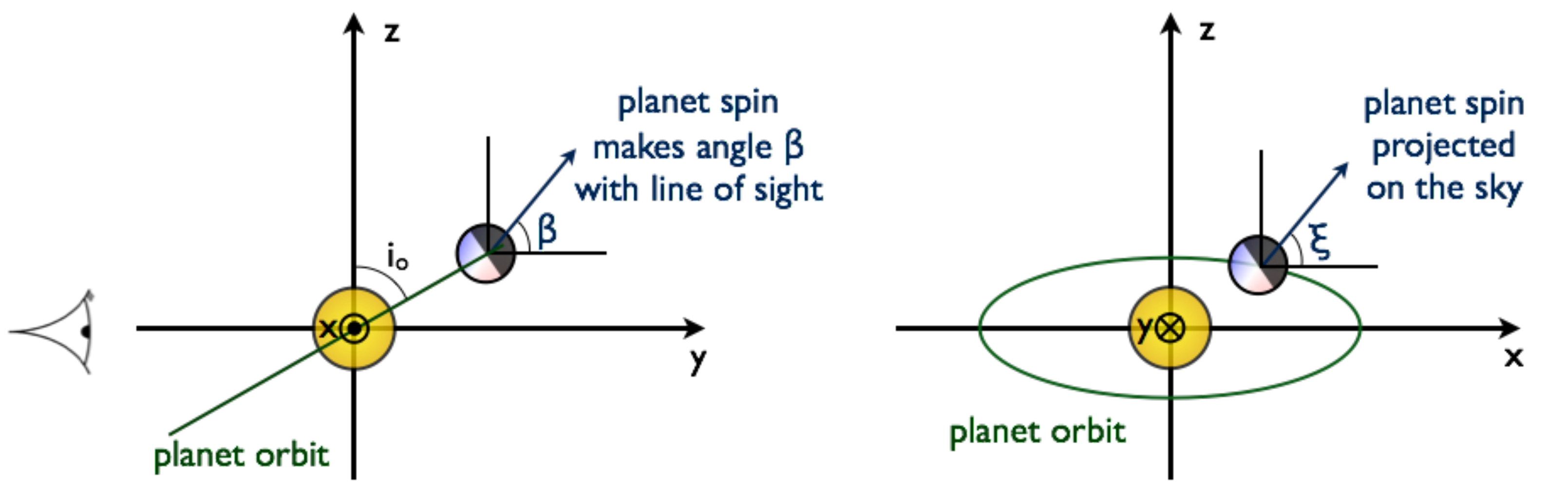}
\caption{Schematic of the sky-projected coordinate system.  The observer is looking along the $\hat{y}$ direction, while the plane of the sky coincides with the $x-z$ plane. {\em Left:} The system viewed from the positive $\hat{x}$ direction. The observer is shown at the left, and the orbit of the planet projected on the $y-z$ plane is a straight line tilted by the inclination angle $i_o$ with respect to the $y$ axis. The angle $\beta$ is the cone angle between the planet's spin and the line of sight (the $\hat{y}$ direction). {\em Right:} The system viewed from the negative $\hat{y}$ direction (the observer's point of view). The orbit of the planet is shown projected on the sky plane, and is represented by a circle that has been tilted by the inclination angle $i_o$. When the spin vector is projected onto the sky plane, its orientation is described by the angle $\xi$ that it makes with respect to the $x$ axis.}
\label{fig:fig2}
\end{figure*}

We assume that the planet/moon system has an orbit that has been circularized and is coplanar with the orbit of its CM around the host star. The velocity of the primary planet around the planet/moon CM is then:
\begin{equation}
V_{p} = -\Omega_{p} a_{p} \left( \frac{m_{\leftmoon}}{m_{p}+m_{\leftmoon}} \right) \left[\cos \phi_p \, \hat{X} + \sin \phi_p  \, \hat{Y} \right] \, 
\end{equation}
where the leading factor outside the square brackets is just the radial velocity semiamplitude of the primary planet in its orbit about the planet/moon CM ($K_p$, see Equation \ref{planetsemiamplitude} in the main paper), and $\phi_p$ is the planet/moon system's orbital phase. The same $X-Y$ orbital plane is used here as for the outer orbit.

If we now perform a rotation about the $\hat{X}$ axis, which lies in the plane of the sky, by an angle $i_o$ (the orbital inclination angle), we find for the radial velocities:	
\begin{equation}
{V_{ro} =  \frac{\Omega_o a_o}{\sqrt{1-e_o^2}}\left[ \sin(\varphi-\varphi_e) - e_o \cos \omega_o \right] \,\sin  i_o }
\end{equation}
\begin{equation}
V_{rp} = \Omega_{p} a_{p} \left( \frac{m_{\leftmoon}}{m_{p}+m_{\leftmoon}} \right)  \sin \phi_p  \, \sin i_o
\end{equation}

All of the above holds whether the primary planet is self-luminous or illuminated by the host star.  The following, however, holds only for externally illuminated planets.  For self-luminous planets, the following terms average to zero (unless some other asymmetry is introduced in the planet's brightness profile, like a disk clump occultation or planetary activity).  Computing the Doppler shifts for the case of an externally illuminated planet is the more challenging part of the calculation. 

We first select a set of unit vectors, $\hat{n}'$, which point from the center to the surface of the planet.  The two angles describing each unit vector are $\Theta$ and $\Phi$ such that
\begin{equation}
\hat{n}' = \sin \Theta \cos \Phi ~\hat{x}' -\cos \Theta ~ \hat{y}'  + \sin \Theta \sin \Phi~\hat{z}'
\end{equation}
(see the diagram in Figure~\ref{fig:fig1} for the definition of the $\{x',y',z'\}$ coordinate system).
In order to populate these vectors over the hemisphere that is illuminated by the very distant host star, the angles $\Phi$ are uniformly sampled around $2 \pi$, while the angle $\Theta$ is distributed with a probability per unit $\Theta$ of $2 \cos \Theta \sin \Theta$.

For different true anomalies of the outer orbit, and an outer orbital inclination angle of $i_o$, a distant observer sees these normal vectors as
\begin{eqnarray}
{n}_x & = & {n}'_x \cos \varphi_o - {n}'_y \sin \varphi_o \\
{n}_y & = & {n}'_x \sin \varphi_o \sin i_o + {n}'_y \cos \varphi_o \sin i_o - {n}'_z \cos i_o  \\
{n}_z & = & {n}'_x \sin \varphi_o \cos i_o + {n}'_y \cos \varphi_o \cos i_o + {n}'_z \sin i_o
\end{eqnarray}
where the directions $\{\hat{x},\hat{y},\hat{z}\}$ refer to the observer's coordinates, and, in particular, $\hat{y}$ is the observer's view direction (see Figure~\ref{fig:fig2} for a schematic of the geometry).  {Here we have used a shorthand notation with $\varphi_o \equiv \varphi -\varphi_e$.}
The velocity of a point on the planet's surface specified by $\hat{n}$ has a velocity in the observer's fixed coordinate system of $\vec{V}_s = r_{p} \,\vec{\omega}_s \times \hat{n}$ which can be written as the expansion of the following determinant:
\begin{center}
$ \vec{V}_s = r_{p}
\left|
\begin{array}{ccc}
 \hat{x} & \hat{y} & \hat{z} \\
\omega_{s,x} & \omega_{s,y} & \omega_{s,z} \\
n_x & n_y & n_z \\
\end{array}
\right| $
\end{center}
where the angular velocity vector, $\vec{\omega}_s$, rotated to the observer's fixed coordinate system has the following components:
\begin{eqnarray}
\omega_{s,x} & = &\omega_s \sin \beta \cos \xi  \\
\omega_{s,y} & = &\omega_s \cos \beta  \\
\omega_{s,z} & = & \omega_s \sin \beta \sin \xi 
\end{eqnarray}
and where $\beta$ is the angle between the planet's spin vector and the observer's view direction, and $\xi$ is the angle that the planet's spin vector, projected onto the sky plane, makes with respect to the $x$ axis (see the geometry in Figure~\ref{fig:fig2}).

In general, the cross product, $\vec{\omega}_s \times \hat{n}$ is a fairly messy expression, except for the trivial case where $\beta = 0$ (i.e., we are looking along the planet's spin vector), in which case the radial component of $\vec{V}_s = 0$.  However, the determinant is simple enough to evaluate numerically.   

The information contained in the $\hat{n}$ vectors is sufficient to determine (1) whether that point on the planet's surface is visible at a given instant, and (2) the surface brightness of that point, depending on the scattering law.  To determine the visibility of the surface point specified by a given $\hat{n}$, we simply examine the dot product of that vector with the line of sight vector $\hat{y}$, and require that it be negative:
\begin{equation}
\hat{y} \cdot \hat{n} = n_y = \sin \Theta \cos \Phi \sin \varphi_o \sin i_o -\cos \Theta \cos \varphi_o \sin i_o - \sin \Theta \sin \Phi \cos i_o < 0
\end{equation}
In the restricted case of viewing the outer orbit edge on, this reduces to the condition that
\begin{equation}
\sin \Theta \cos \Phi \sin \varphi_o  -\cos \Theta \cos \varphi_o  < 0
\end{equation}
\begin{equation}
\tan \Theta \cos \Phi \tan \varphi_o < 1  
\end{equation}

Once the dot product of a normal vector with respect to the view direction has been computed, this yields the cosine of the emission angle with respect to the normal to the planetary surface.  This can be used to compute the intensity of the radiation from that surface element if the scattering law is specified.  For example, if the radiation obeys a Lambertian law, then one multiplies by the dot product to find the relative amount of radiation from that surface element.  For isotropic scattering the factor is just unity, and so forth. Finally, the total integrated radial velocity can be calculated by summing the contribution from each sampled surface element and dividing by the total summed intensity from the samples.


\end{document}